\documentclass[11pt,a4paper]{article}                                
\usepackage[top=1.0in,bottom=1.0in,left=1.0in,right=1.0in]{geometry} 
\usepackage[colorlinks=true,allcolors=blue]{hyperref}                
\usepackage{multicol}                                                
\usepackage{lastpage}                                                
\usepackage{fancyhdr}                                                
\usepackage{hyperref}                                                
\usepackage{pdflscape}  

\usepackage{titlesec}
\titlespacing\section{0pt}{12pt plus 4pt minus 2pt}{0pt plus 2pt minus 2pt}
\titlespacing\subsection{0pt}{12pt plus 4pt minus 2pt}{0pt plus 2pt minus 2pt}
\titlespacing\subsubsection{0pt}{12pt plus 4pt minus 2pt}{0pt plus 2pt minus 2pt}

\usepackage{xurl}

\usepackage{multirow}
\usepackage{multicol}
\usepackage{makecell}

\usepackage{graphicx}   
\usepackage{float}      
\usepackage{subfig}     

\usepackage{listings}                          
\usepackage{amsthm}

\usepackage{amssymb}
\usepackage{siunitx}
\usepackage{chemformula}

\usepackage[english]{babel}
\usepackage{csquotes}
\usepackage[natbib=true,
            style=ieee,
            sorting=none,
            citestyle=numeric-comp
            ]{biblatex}
\usepackage{appendix}

\usepackage{authblk}

\title{Molecular Dynamic Characteristic Temperatures for Predicting Metallic Glass Forming Ability}

\author[a]{Lane E. Schultz}
\author[a]{Benjamin Afflerbach}
\author[a]{Izabela Szlufarska}
\author[a]{Dane Morgan}

\affil[a]{University of Wisconsin-Madison, 1500 engineering Drive, Madison, WI 53706, USA}

\begin{document}
\maketitle

\begin{abstract}
    We explore the use of characteristic temperatures derived from molecular dynamics to predict aspects of metallic Glass Forming Ability (GFA). Temperatures derived from cooling curves of self-diffusion, viscosity, and energy were used as features for machine learning models of GFA. Multiple target and model combinations with these features were explored. First, we use the logarithm of critical casting thickness, $log_{10}(D_{max})$, as the target and trained regression models on 21 compositions. Application of 3-fold cross-validation on the 21 $log_{10}(D_{max})$ alloys showed only weak correlation between the model predictions and the target values. Second, the GFA of alloys were quantified by melt-spinning or suction casting amorphization behavior, with alloys that showed crystalline phases after synthesis classified as Poor GFA and those with pure amorphous phases as Good GFA. Binary GFA classification was then modeled using decision tree-based methods (random forest and gradient boosting models) and were assessed with nested-cross validation. The maximum F1 score for the precision-recall with Good Glass Forming Ability as the positive class was $0.82 \pm 0.01$ for the best model type. We also compared using simple functions of characteristic temperatures as features in place of the temperatures themselves and found no statistically significant difference in predictive abilities. Although the predictive ability of the models developed here are modest, this work demonstrates clearly that one can use molecular dynamics simulations and machine learning to predict metal glass forming ability.
\end{abstract}

{\bf Keywords:} Metallic glasses, Characteristic temperatures, Machine learning, Glass forming ability, Critical casting diameter

\section{Introduction}

\par
Determining metal alloy compositions that have good glass forming ability (GFA), and in particular that yield bulk metallic glasses, has been a grand challenge in metallic glass research for decades. Many previous attempts to predict GFA for metals have made use of some function of the temperatures characterizing important aspects of the melt behavior, sometimes called characteristic temperatures. The characteristic temperatures most used for this purpose are the glass transition ($T_{g}$), the onset of crystallization ($T_{x}$), and liquidus ($T_{l}$) temperatures. For example, the reduced glass transition temperature, $T_{rg}=T_{g}/T_{l}$, is one of the earliest and most iconic GFA indicator \cite{Turnbull1969}. A very successful model for GFA predicts the critical cooling rate, $R_{c}$, as a linear function of $\omega=T_{g}/T_{x}-2T_{g}/(T_{g}+T_{l})$. When fit to 53 metallic glasses, a linear function between $\omega$ and $R_{c}$ had an $R^{2}$ of 0.93 \cite{Long2009}. $T_{rg}$ and $\omega$ are just two of over 20 functions of $T_{g}$, $T_{x}$, and $T_{l}$ that have been proposed to quantify GFA \cite{Deng2020, Xiong2019}. Although successful models have only been shown for a small number of alloys, there are clear indications that insights into GFA can be given by characteristic temperatures.

\par
Another GFA indicator, in this case from the melt, is the liquid fragility, $m$, which is measured by finding the slope of viscosity as a function of temperature near $T_{g}$ for an alloy. Glasses with higher viscosities when approaching $T_{g}$ are said to be strong (low $m$) and are thought to suppress the kinetics for crystallization. Conversely, glasses that experience low viscosities upon cooling are said to be fragile \cite{Dai2019-2, Jaiswal2016}. It was shown that a linear combination of $T_{rg}$ and $m$ fit to $log_{10}(D_{max}^{2})$ for 42 glassy alloys had an outstanding $R^{2}$ score of 0.980 \cite{Johnson2016}. Here, $D_{max}$ is the critical casting diameter. Hence, $D_{max}$ can be written as a relatively simple function of $T_{g}$, $T_{l}$, and $m$. Although $m$ is not strictly a characteristic temperature it is similar in spirit as it represents the temperature dependent physics of the melt, and it can be related to a characteristic temperature as discussed later.

\par
Quantitative relationships between the most important and intrinsic measure of GFA, specifically $R_{c}$ and $D_{max}$, and characteristic temperatures (or closely related melt properties) can be established. However, the aforementioned relationships have limited utility for new material discovery because a glass must be synthesized to obtain the features $T_{g}$, $T_{x}$, and $m$. $T_{l}$ must also be obtained, although this temperature can be measured without making a glass and is often accessible through thermodynamic modeling without actual synthesis. An exciting design opportunity would be realized if we could access $T_{g}$, $T_{x}$, $T_{l}$, and $m$ from molecular simulations, as this would allow the above correlations to be used for computational prediction of GFA. It is this opportunity that motivate the present work.

\par
$T_l$ can be predicted from molecular simulations quite accurately \cite{Hafner1983, Hong2015, Jinnouchi2019} and is often available from thermodynamic models fit to experiments for relevant alloys, so we will not focus further on this quantity and simply take it from experiments or online phase diagrams when available for the rest of the present study. $T_{g}$, $T_x$, and $m$ are all in theory accessible to molecular simulations but have major practical challenges. $T_{g}$ from molecular dynamic (MD) studies are strongly impacted by the very fast cooling rates necessitated by the short time scales accessible to MD. The MD values of $T_{g}$ tend to be higher than experimental values due to heating/cooling rate differences and adopted methodology \cite{Aliaga2019, Musgraves2019}. See Ref.~\cite{Aliaga2019} for a comparison of experimental and MD approaches for finding $T_{g}$. Finally, another set of limitations are imposed by $T_x$ calculation. As seen in Ref.~\cite{Louzguine-Luzgin2020}, the crystallization kinetics in MD for a single composition varies by system size and annealing temperatures explored. The aforementioned indicates that mimicking an experimental $T_x$ value from differential scanning calorimetry (DSC) would pose many challenges in choices of system sizes, cooling rates to attain a glass, heating rates, and the starting temperature to heat a material. $m$ is also difficult to practically calculate from MD as it requires determining viscosity as a function of temperature near $T_{g}$, which is impractical due to the slow kinetics near $T_{g}$. Equations with reliable extrapolation to low temperature viscosities have their own set of challenges for finding their fitting parameters with MD \cite{Blodgett2015, Chen2020}. Due to these obstacles, direct MD of some experimental characteristic temperatures is currently impractical.

\par
Nevertheless, MD accessible quantities that approximate or correlate with some of the previously defined material properties exist. For example, Kelton et al. have shown correlations between $m$ and the ratio between $T_{g}$ and a temperature where a set of compositions cross a set viscosity value, $T^{*}$ \cite{Gangopadhyay2017-2, Gangopadhyay2017-1}. They further showed that $T_{g}$ can be captured as a function of $T^{*}$, and the crossover from Arrhenius behavior temperature, $T_{A}^{*}$ \cite{Dai2019-1}. Specifically, a fit between $T_{A}^{*}/T_{g}$ and $T_{g}/T^{*}$ had an $R^{2}$ of 0.96 \cite{Gangopadhyay2017-1}. The results from Kelton et al. and in Ref.~\cite{Johnson2016} together imply that $D_{max}$ should be a function of $T_{g}$, $T^{*}$, $T_{A}^{*}$, and $T_{l}$.

\par
Other alloy characteristic temperatures are MD acquirable and included in this study. We choose to examine self-diffusion derived temperatures because trends between self-diffusion, viscosity, and relaxation times with respect to temperature show strong relationships, as seen in Ref.~\cite{Puosi2018}. Parallel to the way $T^{*}$ and $T_{A}^{*}$ are defined for viscosity, we define $T^{'}$ and $T_{A}^{'}$ as the temperatures where diffusivity reaches a critical value and where diffusivity deviates from an Arrhenius trend, respectively. We can find an approximate $T_{g}$ by direct high-rate cooling, which we call $T_{gm}$. Approximate $T_{g}$ values can also be estimated by fitting to high-temperature kinetic properties with a simple Vogel-Fulcher-Tamman (VFT) form and extrapolating. We use this approach to define $T_{g}^{*}$ as the temperature where extrapolated viscosity reaches $10^{12}$ $Pa \cdot s$ \cite{Angell1995} and $T_{g}^{'}$ as the temperature where extrapolated self-diffusion values reach $10^{-12}$ $\text{\AA}^{2}/ps$ (which is the method used to define $T_{g}$ in Ref.~\cite{Chen2016}).

\par
We argue and show that GFA insights could be gained from the MD characteristic temperatures of $T_{gm}$, $T_{g}^{*}$, $T^{*}$, $T_{A}^{*}$, $T_{g}^{'}$, $T^{'}$, $T_{A}^{'}$, and $T_{l}$. We use both regression and classification machine learning (ML) models to quantify the ability of models to learn GFA from MD characteristic temperatures. Although the examined data sets were small, preliminary results suggest that MD quantification of GFA is possible.

\section{Methods}

\subsection{Molecular Dynamics}

\par
All MD simulations were performed with the LAMMPS package \cite{Plimpton1995}. The time step was set to 1 $fs$. Periodic boundary conditions were applied along all directions. Finnis-Sinclair (FS) and embedded-atom model (EAM) potentials were used to simulate alloys of interest \cite{Becker2013, Cheng2009, Cheng2008, Fujita2010, Hale2018, Li2019, Sheng2011}. The starting MD supercell structures were built with a repeating simple cubic unit cell with 1,000 atoms. The element type for each atom was assigned randomly to match compositions of interest. An isothermal hold at 2,000 K was run for 100 $ps$ under the NPT ensemble and was used as the initial trajectory for melt quench simulations. In the NPT ensemble, N is the number of atoms, P is the pressure, and T is the temperature and are all held constant.

\par
Continuing from the starting structures, isothermal holds were constructed by dropping 100 $K$ from the preceding hold. Each of the holds were ran for 10 $ps$ which gave a cooling rate of $10^{13}$ K/s. The final temperature probed was 100 $K$. For each of the isothermal holds from the melt-quench simulation, the final trajectory was run for an additional 10 $ns$ under the NVT ensemble. In the NVT ensemble, N is the number of atoms, V is the system volume, and T is the temperature and are all held constant. For self-diffusion and viscosity, the reference time for calculations was at the beginning of the 10 $ns$ isothermal hold. Mean squared displacement (MSD) for self-diffusion, viscosity, and averaged thermodynamic data were attained from the last 2 $ns$ of the 10 $ns$ isothermal hold. Each composition was run 2 times with different starting atomic positions for averaging characteristic temperature measurements to reduce uncertainty.

\subsection{Kinetic Properties}

\par
Self-diffusion for each isothermal hold from quench runs was calculated through the long-time limit of MSD with Equation~\ref{diffusion} \cite{Rapaport}. In Equation~\ref{diffusion}, $D$ is the self-diffusion, $N$ is the total number of atoms, $t$ is the time, and $r$ is the position of an atom $i$. If the average mean squared displacement of atoms is less than 1 $\text{\AA}^{2}$, we assumed that the atoms cannot be reliably identified as diffusing rather than just vibrating in place and the associated self-diffusion value was excluded from all future analysis and our VFT fits to data.

\begin{equation}\label{diffusion}
    D=\lim_{t \to \infty} \frac{1}{6Nt} \left < \sum_{i=1}^{N} \left [ r_{i}(t)-r_{i}(t=0) \right ]^{2} \right >
\end{equation}

\par
The Green-Kubo formalism was used to calculate viscosity (Equation~\ref{gbvisc}) in an equilibrated system by integrating the autocorrelation of the pressure tensor off-diagonals \cite{Rapaport, Puosi2018, Hess2002, Zwanzig1964}. In Equation~\ref{gbvisc}, $k_{B}$ is the Boltzmann's constant, $T$ is the temperature, $V$ is the system volume, $t_{0}$ is the starting time, $t$ is a time value, and $P_{ij} \in \{P_{xy}, P_{xz}, P_{yz} \}$ are the elements of the pressure tensor. For a three-dimensional simulation, the integral of the autocorrelation of $P_{xy}$, $P_{xz}$, and $P_{yz}$ can be averaged together due to their symmetry equivalence in the liquid state.

\begin{equation}\label{gbvisc}
    \eta = \frac{V}{k_{B}T} \int_{0}^{\infty} \left < P_{ij}(t_{0})P_{ij}(t_{0}+t)\right >dt 
\end{equation}

\par
Although several expressions exist to fit dynamic properties for a fluid, we find that the VFT expression was the simplest to fit and use and we did not see any advantage using more complex functional forms \cite{Blodgett2015}. The VFT function was used to fit the resulting self-diffusion and viscosity data (Equation~\ref{vft}). In the VFT expression, $A$ , $B$, and $T_{0}$ are fitting constants. $x$ is either self-diffusion or viscosity depending on which data was used for fitting.

\begin{equation}\label{vft}
    log_{10}(x)=A+\frac{B}{T-T_{0}}
\end{equation}

\subsection{Characteristics Temperatures}\label{cts}

\par
\begin{figure}[H]
\centering
\includegraphics[width=0.95\textwidth]{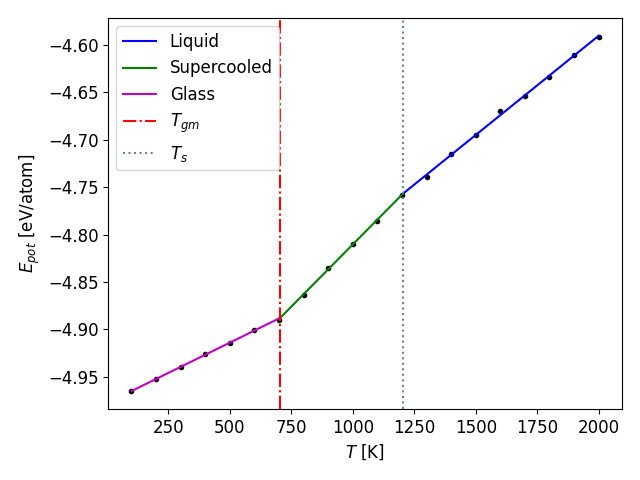}
\caption{The results of a three part piecewise linear fit to the potential energy vs. temperature for $Cu_{50}Zr_{50}$. The transition from low-temperature glassy to mid-temperature supercooled liquid regimes is shown by the red vertical line which denotes the molecular dynamic glass transition temperature for a single run.}
\label{single_tg}
\end{figure}

Seven temperatures were calculated from MD: $T_{gm}$, $T_{g}^{*}$, $T^{*}$, $T_{A}^{*}$, $T_{g}^{'}$, $T^{'}$, and $T_{A}^{'}$. $T_{l}$ was also considered but was taken from experimental data or the Alloy Phase Diagram Database from the American Society for Metals (ASM) International. Table~\ref{asm} lists the system and the corresponding diagram used from ASM for $T_{l}$ values. Here we describe how each characteristic temperature was determined. $T_{gm}$ was calculated via methods used in Refs. \cite{Cheng2008_2, Sheng2012} which use a change in the potential energy, $E_{pot}$, slope between high and low temperature regimes to determine $T_{gm}$. The change of slope is typically acquired through a ``knee" in a heat capacity curve with respect to temperature but we did not use this approach due to issues with numerical noise. We find the change in slope as follows. We perform a fit to $E_{pot}$ vs. temperature using three piecewise linear fits, which represent the behavior of the system in the liquid, supercooled, and glassy phases. The three fitted lines are chosen to minimize the squared residuals using the package in Ref.~\cite{pwlf}. An example of the fitting is shown in Figure~\ref{single_tg}. We then determine $T_{gm}$ as the intersection of the two lower temperature lines (the glassy and supercooled liquid lines). We note that the intersection of the two higher temperature lines generally gives a dynamical slow down temperature, $T_{s}$, as defined in Ref.~\cite{Sheng2012}. We were not able to determine a robust $T_{s}$ for all systems and we did not use this value as a characteristic temperature in this study. We used this approach to determine $T_{gm}$ for all 95 compositions studied. The uncertainty in our estimate of $T_{gm}$ is found from the standard error of the mean (SEM) across the 2 cooling runs. The average SEM for $T_{gm}$ across multiple compositions were 19 $K$, which is adequately low given the other uncertainties in this overall analysis.

\par
\begin{figure}[H]
\centering
\includegraphics[width=0.95\textwidth]{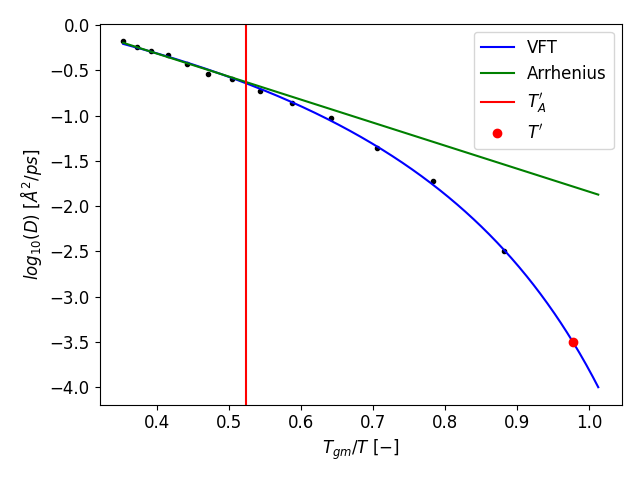}
\caption{The self-diffusion behavior with respect to temperature for a single run of $Cu_{50}Zr_{50}$. The black points are the molecular dynamic self-diffusion for NVT isothermal holds. The blue and green curves are the VFT and high temperature Arrhenius fit to self-diffusion data respectively. The red point denotes the temperature at a user specified self-diffusion cutoff and the vertical line represent the temperature where self-diffusion deviates from Arrhenius behavior.}
\label{tprime}
\end{figure}
\begin{figure}[H]
\centering
\includegraphics[width=0.95\textwidth]{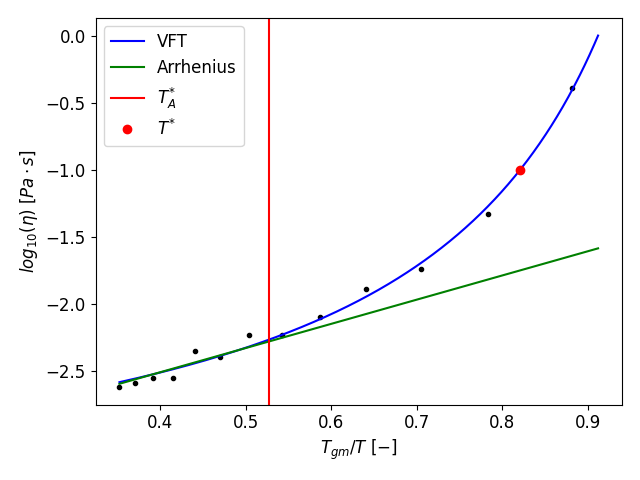}
\caption{The viscosity behavior with respect to temperature for a single run of $Cu_{50}Zr_{50}$. The black points are the molecular dynamic viscosity for NVT isothermal holds. The blue and green curves are the VFT and high temperature Arrhenius fit to viscosity data respectively. The red point denotes the temperature at a user specified viscosity cutoff and the vertical line represent the temperature where viscosity deviates from Arrhenius behavior.}
\label{tstar}
\end{figure}

Self-diffusion and viscosity follow an Arrhenius behavior at high temperatures. At lower temperatures, cooperative motion becomes more significant leading to a deviation from Arrhenius behavior \cite{Iwashita2013, Jaiswal2016}. A simple algorithm was applied to consistently compute $T_{A}^{'}$ and $T_{A}^{*}$ across multiple compositions. First, self-diffusion and viscosity values were fit to VFT and a linear function. We then compared the two fits to see where they deviated and used that measure to determine the characteristic temperature. More specifically, we first started with all the data points. Then, if the mean absolute residuals (MAR) of the linear fit were more than the VFT fit by an amount greater than 0.005 for either self-diffusion or viscosity in their respective $log_{10}$ units, the lowest temperature points were excluded from linear fits until the linear minus VFT MAR equaled or fell below the threshold. The computed $T_{A}^{'}$ and $T_{A}^{*}$ denote the lowest temperatures where VFT and Arrhenius fits are approximately indistinguishable. The present approach provides a consistent definition for $T_{A}^{'}$ and $T_{A}^{*}$ across the 95 studied compositions. The uncertainty in our estimate of these characteristic temperatures is found similarly to $T_{gm}$ above from SEM across the 2 cooling runs. The algorithm used to compute $T_{A}^{'}$ and $T_{A}^{*}$ have average SEM values of 21 $K$ and 11 $K$ respectively for all compositions studied. More averaging could reduce uncertainties, but we find the uncertainties sufficient for the current work. A sample calculation of $T_{A}^{'}$ and $T_{A}^{*}$ are shown in Figures~\ref{tprime} and \ref{tstar} respectively.

\par
In a similar manner to $T_{A}^{'}$ and $T_{A}^{*}$ calculation, $T^{'}$ and $T^{*}$ were determined using fits to the VFT equation of self-diffusion and viscosity data respectively. $T^{'}$ was defined here as the temperature at which each composition reached a self-diffusion value of $10^{-3.5}$ $\text{\AA}/ps$ (Figure~\ref{tprime}). Similarly, $T^{*}$ was defined as the temperature where each composition reached a viscosity value of $10^{-1}$ $Pa \cdot s$ (Figure~\ref{tstar}). The choice of cutoffs were as large as practical given a goal of being near data points acquired through MD to reduce extrapolation error from VFT. Using the same error methods for the characteristic temperatures described above gave average SEM for $T^{'}$ and $T^{*}$ of 18 $K$ and 9 $K$ respectively. The values of $T_{g}^{'}$ and $T_{g}^{*}$ were also determined from VFT fits to self-diffusion and viscosity. Again, using the same error methods, when the VFT functions were extrapolated to $10^{-12}$ $\text{\AA}^{2}/ps$ and $10^{12}$ $Pa \cdot s$ for self-diffusion and viscosity, respectively, the average SEM for $T_{g}^{'}$ and $T_{g}^{*}$ were 35 $K$ and 31 $K$, respectively. Each CT along with its description can be seen in Table~\ref{CTs_def}.

\begin{table}[H]
\centering
\caption{The definitions for all CTs included in this study.}
\begin{tabular}{|l|p{10cm}|}
\hline
CT                                & Description                                                                                                           \\ \hline
\hline
$T_{gm}$                          & The glass transition temperature acquired from a potential energy versus temperature relationship.                    \\ \hline
$T_{g}^{*}$                       & The glass transition temperature acquired by extrapolating VFT viscosity to $10^{12}$ $Pa \cdot s$.              \\ \hline
$T^{*}$                           & The temperature where viscosity reaches $10^{-1}$ $Pa \cdot s$.                                                    \\ \hline
$T_{A}^{*}$                       & The deviation from high temperature Arrhenius behavior in a viscosity versus temperature relationship.                \\ \hline
$T_{g}^{'}$                       & The glass transition temperature acquired by extrapolating VFT self-diffusion to $10^{-12}$ $\text{\AA}^{2}/ps$. \\ \hline
$T^{'}$                           & The temperature where self-diffusion reaches $10^{-3.5}$ $\text{\AA}/ps$.                                          \\ \hline
$T_{A}^{'}$                       & The deviation from high temperature Arrhenius behavior in a self-diffusion versus temperature relationship.           \\ \hline
$T_{l}$                           & The liquidus temperature.                                                                                             \\ \hline
\end{tabular}
\label{CTs_def}
\end{table}

\subsection{Data}\label{sect_data}

\par
Experimental $T_{l}$, $D_{max}$, and melt-spun classification data were acquired through an online database with citations to papers included in the Data Availability section. Missing values of $T_{l}$ were replaced by values read from ASM phase diagrams (see Table~\ref{asm}). Although $T_{l}$ did not come from MD, previous studies show that MD calculation of $T_{l}$ is possible, as discussed in the Introduction. When multiple $D_{max}$ or $T_{l}$ values were available for the same composition, the mean of the values were used. If any one of the melt-spun sheets resulted in fully or partially crystalline sheets, then that alloy was classified as a Poor GFA alloy. Fully amorphous sheets and alloys that had a $D_{max}$ measure were classified as Good GFA alloys. As noted above, the data set containing only $D_{max}$ data has 21 compositions, is a subset of the 95 total alloys, and is called the regression data set. The full set of 95 compositions is called the classification data and contains classes of Poor GFA (39 compositions = 41\% of the data) and Good GFA (56 compositions = 59\% of the data).

\par
For our classification data, we can generate simple new features from products, ratios, summations, and differences (PRSDs) of the characteristic temperatures, which we call PRSD features. We can then compare the effect of the classification models using the PRSD features to models using our original set of characteristic temperatures as features. We are motivated to explore the PRSD features by the long history of studies using linear functions of PRSD features to predict different aspects of GFA, as discused in the Introduction. To construct the PRSD features, we follow the approach of Ref.~\cite{SCHULTZ2021110494}. We start with a set of characteristic temperatures defined as $A=\{T_{gm}, T_{g}^{*}, T^{*}, T_{A}^{*}, T_{g}^{'}, T^{'}, T_{A}^{'}, T_{l}\}$. From set $A$, summations and differences were taken between each of the features to construct set $B$. A sample feature contained in set $B$ would be $(T_{gm}-T_{l})$. Now define $C=A \cup B$. From $C$, we can take powers up to $n$ for every element to produce set $D$. For instance, $(T_{gm}-T_{l})^{2}$ is an element in set $D$. For our current work, we limited $n \in \{1,2\}$. Define set E as follows: $E=C \cup D$. For every element in E, we can take the inverse to produce set $F$. Continuing from our example, a possible element produced would be $1/(T_{gm}-T_{l})^{2}$. We then construct another set $G=E \cup F$. The final operation to generate features involves products between every combination of two elements from set $G$ to produce set $H$. The final feature set was defined as $X=G \cup H$. This process produced a total of 32,896 features. As noted in the Results and Discussion section, PRSD features do not provide statistically significant improvements in learning GFA compared to using CTs alone. Hence, a larger space of features was not explored (e.g., $n \in \{1,2,3,4,...\}$ for power features). All data sets used for machine learning, models, and figures can be found at figshare and GitHub at Refs.~\cite{figshare_data} and \cite{github_data} respectively.

\subsection{Machine Learning}\label{sect_ml}

\par
Scikit-learn was used for ML applications \cite{scikit-learn}. The XGBoost model type was attained from Ref.~\cite{xgboost}. A standard scaler was used to transform all our features to have a mean of zero and a standard deviation of 1 for each dataset. When used in nested CV, the scaler is trained only on the training set and then used to transform both the training and test features.

\par
Since the number of fitting points for our regression data are small, we fit a simple model to our characteristic temperatures of greatest importance. We use $log_{10}(D_{max})$ as the target feature. Raising $D_{max}$ to a power within a logarithm like in Ref.~\cite{Johnson2016} has no impact on the fitting ($log_{10}(D_{max}^{x}) = xlog_{10}(D_{max})$ and ML models can account for the multiple by a real number $x$) so no power is included. First, we trained a Least Absolute Shrinkage and Selection Operator (LASSO) model that minimized root mean squared error ($RMSE$) through a grid search of $\alpha$ hyperparameter values \cite{lasso}. The $\alpha$ values considered were $10^{-5}$ to $10^{5}$ in a $log_{10}$ grid of 100 values. The absolute value of weights from the fitted model denote the magnitude of the model's response with respect to a change in feature value which is a measure of the contribution each feature makes to predicting $log_{10}(D_{max})$. Using the two features with the highest weights, an OLS regression model was fit with an intercept term to produce the final model (using more features led to overfitting and reduced the CV accuracy, as might be expected given the complex physics and limited training data in the model). To assess the OLS model, 20 repeats of 3-fold CV were performed to view the effects of prediction on data outside of the training set. Metrics reported for final regression models are the mean average error ($MAE$), coefficient of determination ($R^{2}$), $RMSE$, and the $RMSE/\sigma$ where $\sigma$ denotes the spread on predicted target values.

\par
We implemented repeated nested CV to assess classification models. Model types included are Gradient Boosting (GB), eXtreme Gradient Boosting (XGBoost), and Random Forest (RF) which are ensemble models \cite{Pavlov2019, Friedman2001, xgboost}. We used ensemble models because of their tendency to outperform linear models in $D_{max}$ predictions in our past research \cite{SCHULTZ2021110494}. Models were trained on PRSD feature and original characteristic temperature feature sets. Data were shuffled and split into 3 outer and 3 inner folds. The choice of hyperparameters that minimized $RMSE$ were determined from the inner folds via a grid search (Table~\ref{gridsearch}). The outer folds were used to assess a model's performance on data not used for training. Nested CV was performed 20 times which gave 60 test sets for each model and data set combination. Classification scores were averaged between all leave out sets. A two-sided T-test was performed between models trained on the original versus the PRSD feature sets to show if there was a statistically significant difference between training on the two feature sets.

\begin{table}[H]
\centering
\caption{The grid of hyperparameters for XGBoost, GB, and RF models. The conventions of Scikit-learn and XGBoost were used for parameter names \cite{scikit-learn, xgboost}.}
\begin{tabular}{|ccc|}

\hline
{Model}                 & {Parameter}   & {Values}                                \\
\hline
\hline

\multirow{3}{*}{RF}   & n\_estimators       & 30, 40, 50, 60, 100, 500            \\
                               & max\_features & sqrt, log2, None                 \\
                               & max\_depth              & 2, 3, 4, None          \\
                               
\hline

\multirow{4}{*}{GB}   & learning\_rate              & 0.001, 0.01, 0.1, 0.2       \\
                               & n\_estimators       & 30, 40, 50, 60, 100, 500   \\
                               & max\_features & sqrt, log2, None                 \\
                               & max\_depth              & 2, 3, 4                \\
\hline

\multirow{4}{*}{XGBoost}   & learning\_rate              & 0.001, 0.01, 0.1, None       \\
                           & max\_depth              & 2, 3, 4, 5, None                \\
                           & subsample              & 0.5, 0.8, 1.0, None                \\
                           & gamma              & 0, 1, 5, None                \\

\hline

\end{tabular}
\label{gridsearch}
\end{table}

\par
All classification models were assessed with the area under the curve (AUC) from Precision-Recall (PR) curves and maximum $F1$ scores. $F1$ is defined as the harmonic mean between precision and recall. The baseline AUC for any PR curve is defined as $P/(P+N)$ with $P$ being the number of positive and $N$ being the number of negative cases \cite{Saito2015}. For the nested CV tests, PR curves where averaged together for the outer loops. The averaging was performed by first building a grid of horizontal values (recall for PR) from 0 to 1 with 1,000 linearly spaced values. Then data are linearly interpolated. This ensured that all averaged values were gridded equally for vertical averaging. For $F1$ scores, the maximum values were averaged between all outer folds.

\par
SHapley Additive exPlanations (SHAP) provide feature interpretations by fairly allocating feature contributions via game theory, an approach developed by Professor Lloyd Shapley \cite{Strumbelj2014}. SHAP values were attained with the shap package in Ref.~\cite{shap} and used to analyze feature contributions for an XGBoost model trained on all 95 cases (called the full-fit model) of our classification data set. We will show that the prediction contributions of the top 3 features agree with physical intuition of GFA.

\section{Results and Discussion}

\par
\begin{figure}[H]
\centering
\subfloat[Predicted Back]{
\label{parity_all}
  \includegraphics[clip,width=0.7\columnwidth]{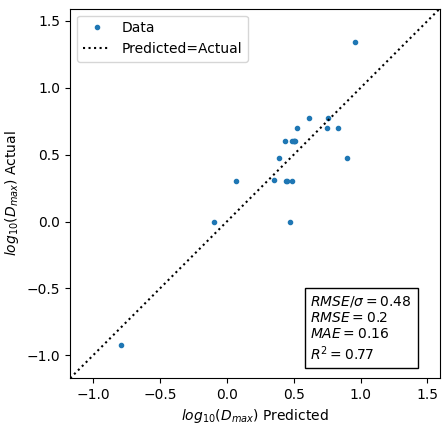}
}
\\~\\
\subfloat[CV Averaged]{
\label{parity_cv}
  \includegraphics[clip,width=0.7\columnwidth]{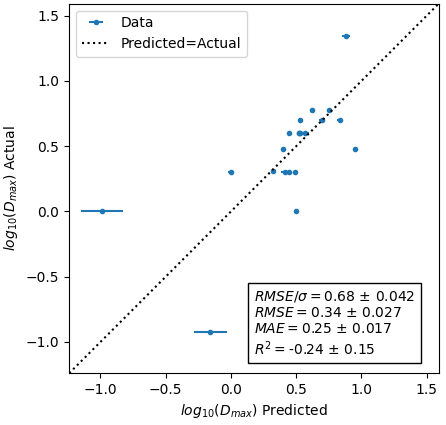}
}
\caption{The parity plots for OLS models along with standard ML performance metrics. Each of the blue points denotes a prediction of $D_{max}$ from a model. The black dotted line represents where ideal predictions would fall. We note a reduction in prediction ability of OLS models when a cross validation test was performed (predicted back compared to CV averaged).}
\label{rc_lasso}
\end{figure}

First, we consider the regression data set and models. The OLS regression model fit from LASSO selected features was used to predict back onto the training set to produce the parity plot in Figure~\ref{parity_all}. The $RMSE/\sigma$ was 0.48 which means that the $RMSE$ of predicted $log_{10}(D_{max})$ values are well-below the spread in true values, $\sigma$. The characteristic temperatures with the highest absolute weights and therefore used in the model were $T_{l}$ and $T_{A}^{'}$. The sign of weights for $T_{l}$ and $T_{A}^{'}$ were negative and positive respectively which follow expected theories. Assuming the experimental $T_{g}$ is similar across studied compositions, a higher $T_{l}$ represents a larger range of temperatures through which the supercooled liquid must remain stable without nucleating crystalline phases before producing a metallic glass. This argument is similar in spirit to that supporting $T_{rg}$ as correlating with GFA, as proposed by Turnbull \cite{Turnbull1969}. Conversely, a higher $T_{A}^{'}$ denotes a dynamic slowdown in a system at a higher temperature, which suppresses the ability of atoms to arrange themselves into an ordered structure and therefore suppresses crystallization and stabilizes glass formation. Higher $T_{l}$ values should therefore reduce GFA while higher values of $T_{A}^{'}$ should increase GFA.

\par
The closest model in literature to our OLS regression model was proposed in Ref.~\cite{Johnson2016} as a linear combination of $T_{rg}$ and $m$. Our model qualitatively agrees with the model in Ref.~\cite{Johnson2016} in two ways. First, lower $T_{l}$ for the OLS regression model generally results in higher $T_{rg}$ and better GFA. Second, $m$ denotes the viscosity of a system as it approaches experimental $T_{g}$. A larger $m$ corresponds to less resistance to movement when cooling and vice versa which is similar to $T_{A}^{'}$. Although the parameters across models are different, they both describe the degree of undercooling along with the mobility of atoms with respect to GFA. The model in Ref.~\cite{Johnson2016} had an $R^{2}$ score of 0.980 while we had an $R^{2}$ score of 0.77. Our scores may have been worse due to the use of different variables, but perhaps also just because of the smaller number of compositions we could study along with our mixing of MD with approximate potentials and experimental data. Some of these aspects could be remedied in future work and potentially approach the outstanding accuracy of the model from Ref.~\cite{Johnson2016} while still using MD derived features.

\par
A more rigorous test of our OLS regression model was performed. The average of predicted values from leave-out compositions in 20 repeats of 3-fold CV along with their SEM is shown Figure~\ref{parity_cv}. As expected, the $RMSE/\sigma$ increased, with values changing from 0.48 to $0.68 \pm 0.042$. The decrease in prediction performance can be explained by the generally complex dependence that might be expected for $log_{10}(D_{max})$ on the features, and in part by the lack of cases with low $D_{max}$ values. The models fit only on high $D_{max}$ cases will fail to predict the cases on the lower extreme. It is not surprising that fitting to few, unevenly distributed data results in a relatively inaccurate (Figure~\ref{parity_cv}) model, even when it may have first appeared promising during the full-fit without cross validation (Figure~\ref{parity_all}).

\par
\begin{figure}[H]
\centering
\subfloat[Original Classification Set]{
\label{Glass_pr}
  \includegraphics[scale=0.75]{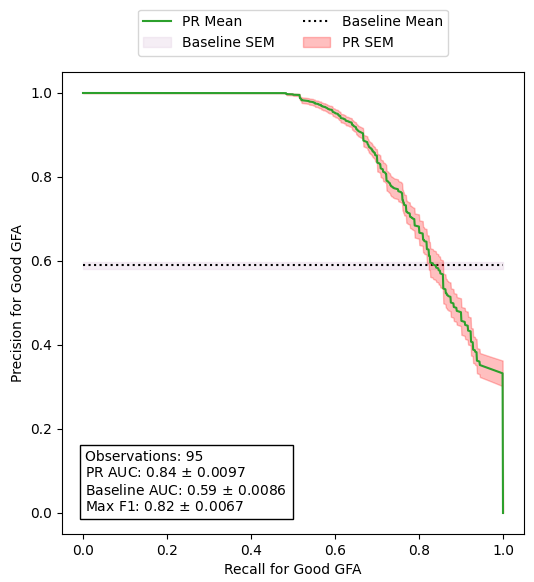}
}
\\~\\
\subfloat[PRSD Classification Set]{
\label{Glass_pr_rf_PRSD}
  \includegraphics[scale=0.75]{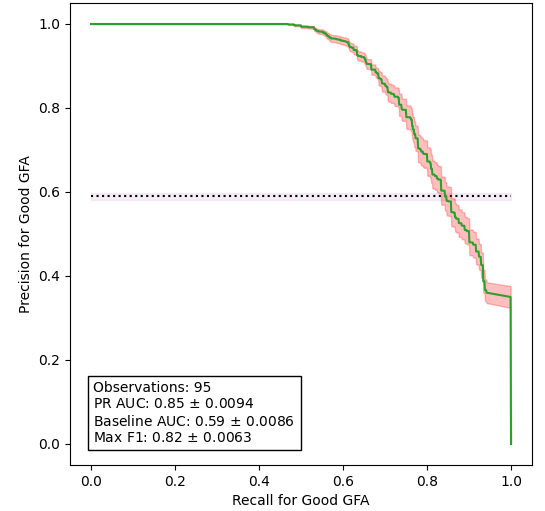}
}
\caption{The comparison between XGBoost models trained on the original versus PRSD classification data sets. The green curve represents the precision and recall given a classification threshold averaged across outer fold test sets. The shaded red area is the SEM between averaged sets. The horizontal line is the average baseline from class counts from outer fold test sets along with the purple shaded region SEM.}
\end{figure}

Second, we consider the classification data set and models. Through nested CV with XGBoost, RF, and GB models fit on the original classification set and then the PRSD classification set, PR curves were produced as detailed in Sec.~\ref{sect_ml} and the scores were tabulated in Table~\ref{ml_scores}. The use of XGBoost models yielded slightly better results than other models and is shown in Figure~\ref{Glass_pr}. The AUC score for PR with Good GFA as positive was $0.84 \pm 0.0097$ with a baseline AUC of $0.59 \pm 0.0086$. The average maximum $F1$ score was $0.82 \pm 0.0067$. One way to understand the implications of the PR curve (Figure~\ref{Glass_pr}) is that if one starts with a list of compositions similar to the training data, and one can accept finding just half the good GFA alloys in the list (50\% recall), then one can be almost certain that the compositions one predicts as good GFAs are correctly identified (100\% precision). As an example, consider searching a list of 100 compositions for GFA with 50 favorable glass formers. The XGBoost model might be expected to predict 25 of the alloys as good glass formers. Almost all 25 would likely be correct, but the model would find only 25 out of the 50 glass formers. A researcher could move along the PR curve to define an acceptable threshold to tune for the number of missed good glass formers while ensuring that a tolerable fraction of glasses studied will produce a glass.

\par
Although the above results are encouraging, they are almost certainly optimistic when using the present model for predicting GFA of general alloys. First, one needs to consider the compositions in the data. Simulations were performed on only 17 chemical systems with a varying number of compositions in each system but 5-6 examples from each system on average. The outer folds used for model assessment therefore most likely contain compositions close to, although not exactly the same as, those used for model training. This will bias the CV scores to be much better than expected on data for totally new systems. Furthermore, new compositions outside the 17 studied here may be quite different in their underlying mechanisms, further reducing the applicability of the model. Finally, it should be noted that the present data is fairly well balanced, with about 59\% of the data with good GFA. Even if the model accuracy is fairly well represented by the present PR on a new test data set, if the fraction of good GFA alloys is much lower, then the model will have much lower precision than found here. We therefore believe that the PR curve obtained here is exciting as it is evidence that GFA can be predicted from MD features based on characteristic temperatures, but it is not a robust guide for expected results on a general screening of alloys for GFA.

\begin{table}[H]
\centering
\caption{The classification scores for all model types. The PRSD classification data contains generated features from the classification set as outlined in Sec.~\ref{sect_ml}.}
\begin{tabular}{|lllrr|}
\hline
{Model} & {Feature Set} & {Metric}                   & {Score} & {SEM} \\ \hline
\hline
GB                  & Classification       & Precision Recall AUC for Good GFA & 0.82           & 0.0086                \\ \hline
GB                  & Classification       & Max F1 for Good GFA               & 0.80           & 0.0061                \\ \hline
RF                  & Classification       & Precision Recall AUC for Good GFA & 0.84           & 0.0078                \\ \hline
RF                  & Classification       & Max F1 for Good GFA               & 0.81           & 0.0059                \\ \hline
XGBoost             & Classification       & Precision Recall AUC for Good GFA & 0.84           & 0.0097                \\ \hline
XGBoost             & Classification       & Max F1 for Good GFA               & 0.82           & 0.0067                \\ \hline
GB                  & PRSD Classification  & Precision Recall AUC for Good GFA & 0.84           & 0.0095                \\ \hline
GB                  & PRSD Classification  & Max F1 for Good GFA               & 0.82           & 0.0065                \\ \hline
RF                  & PRSD Classification  & Precision Recall AUC for Good GFA & 0.86           & 0.0096                \\ \hline
RF                  & PRSD Classification  & Max F1 for Good GFA               & 0.83           & 0.0069                \\ \hline
XGBoost             & PRSD Classification  & Precision Recall AUC for Good GFA & 0.85           & 0.0094                \\ \hline
XGBoost             & PRSD Classification  & Max F1 for Good GFA               & 0.82           & 0.0063                \\ \hline
\end{tabular}
\label{ml_scores}
\end{table}

\par
The use of the PRSD features to fit models (Figure~\ref{Glass_pr_rf_PRSD}) showed some improvement over the original classification feature set. Although initially promising, p-values computed from a two-sided T-test show that there was no statistically significant difference between learning from PRSD and original classification feature sets from 60 test observations. None of the p-values (Table~\ref{pvals}) fell below 0.05 which is a commonly used cutoff for statistical significance. Because of the lesser complexity of the models fit with the original characteristic temperature feature set, the final evaluation metrics of XGBoost without PRSD features were reported.

\begin{table}[H]
\centering
\caption{The p-values from two-sided T-tests.}
\begin{tabular}{|llr|}
\hline
\multicolumn{1}{|c}{Model} & \multicolumn{1}{c}{Metric}       & \multicolumn{1}{c|}{P-Value} \\ \hline
\hline
GB         & Precision Recall AUC for Good GFA & 0.13    \\ \hline
GB         & Max F1 for Good GFA               & 0.22    \\ \hline
RF         & Precision Recall AUC for Good GFA & 0.22    \\ \hline
RF         & Max F1 for Good GFA               & 0.21    \\ \hline
XGBoost    & Precision Recall AUC for Good GFA & 0.59    \\ \hline
XGBoost    & Max F1 for Good GFA               & 0.51    \\ \hline
\end{tabular}
\label{pvals}
\end{table}

\par
\begin{figure}[H]
\centering
\includegraphics[width=0.95\textwidth]{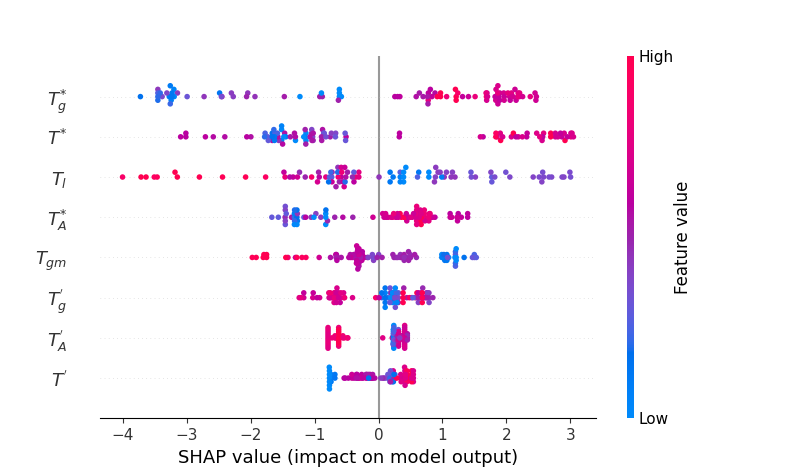}
\caption{The SHAP values for our 8 characteristic temperatures for an XGBoost model. SHAP values from the figure denote the impact from each feature on the prediction of the model. Highest ranked features are displayed from the top to the bottom of the visual.}
\label{shap}
\end{figure}

The SHAP values for our full-fit XGBoost model are shown on Figure~\ref{shap}. The features that have the highest impact on the final prediction are at the top. Conversely, the bottom features have the lowest contribution on final predictions. Feature values to the right of the vertical line in Figure~\ref{shap} push the final prediction to Good GFA while values to the left contribute to Poor GFA classification. The color of values denotes the scale of the feature values. The SHAP values are generally consistent with physical intuition, as can be seen by considering the trends of $T^{*}$, $T_{g}^{*}$, and $T_{l}$. The SHAP values show that higher values of $T^{*}$ and $T_{g}^{*}$ and lower values of $T_{l}$ generally correlate with better GFA. Materials that have higher values of $T^{*}$ experience higher viscosities at higher temperatures. These higher viscosities may slow down the kinetics of crystallization when cooling a molten alloy, supporting better GFA. However, lower values of $T_{l}$ might decrease the range an alloy must cool through before vitrification in a time temperature transformation (TTT) diagram \cite{Musgraves2019}, also supporting better GFA. Similarly, higher values of $T_{g}^{*}$ could narrow the amorphous cooling range on a TTT diagram, again supporting better GFA. These correlations between the SHAP values and these physically sound trends suggest that the model has captured some of the underlying physics behind GFA.

\section{Summary and Conclusion}

\par
We used MD to calculate a set of characteristic temperatures for 95 metal alloys which were used to fit GFA for cases containing $D_{max}$ and melt-spinning data. When using $log_{10}(D_{max})$ as our target, we find a 3-fold cross-validation $RMSE$ score of $0.34 \pm 0.027$ for an OLS regression model, which was not significantly below the standard deviation of 0.43 for the 21 training cases. Only $T_{l}$ and $T_{A}^{'}$ were used in the OLS model. Through nested CV, we assessed the capacity to learn poor from good glass formers for several models. XGBoost was slightly better than all other classification model types. Our average maximum $F1$ score for our RF classification predictions was $0.82 \pm 0.0067$. Additionally, the AUC for classifying Good GFA on our PR curve was $0.84 \pm 0.0097$ which was greater than the baseline of $0.59 \pm 0.0086$ by 0.25. Our XGBoost models predict significantly higher than random guessing. We also determined that learning from PRSD features had no statistically significant effect for any classification tasks compared to learning on the characteristic temperatures themselves. Classification scores suggest that characteristic temperatures from MD can be used as features in machine learning models with predictive ability for GFA. This result provides a potential pathway to discovering new metallic glass alloys based on only simulations. However, to support such screening it is necessary to develop a larger training database that can train more quantitative models with larger ranges of chemistry in their domains of applicability.

\section*{Data Availability}

\par
The raw and processed data required to reproduce these findings are available to download from figshare at \url{https://doi.org/10.6084/m9.figshare.14502135.v1} and GitHub at \url{https://github.com/leschultz/Molecular-Dynamic-Characteristic-Temperatures-for-Predicting-Metallic-Glass-Forming-Ability}.

\section*{Acknowledgements}

\par
Lane E. Schultz is grateful for the Bridge to the Doctorate: Wisconsin Louis Stokes Alliance for Minority Participation, National Science Foundation (NSF) award number HRD-1612530 as well as the University of Wisconsin– Madison Graduate Engineering Research Scholars (GERS) fellowship program for the financial support for graduate student investigation. Other authors gratefully acknowledge support from the NSF Designing Materials to Revolutionize and Engineer our Future (DMREF) program, Division of Materials Research (DMR), METAL \& METALLIC NANOSTRUCTUREs, award number \#1728933. All simulation and machine learning were performed with the computational resources provided by the Extreme Science and Engineering Discovery Environment (XSEDE), National Science Foundation Grant No. OCI-1053575.

\section*{Appendix A}

\setcounter{equation}{0}
\renewcommand{\theequation}{A\arabic{equation}}

\setcounter{table}{0}
\renewcommand{\thetable}{A\arabic{table}}

\setcounter{figure}{0}
\renewcommand{\thefigure}{A\arabic{figure}}

\par
We explain in detail the viscosity calculations used in this work. Because practical MD has a finite run time, the integral form Equation~\ref{gbvisc} has an upper time limit of $t_{s}$. MD is also discrete so integration was represented with summation instead. Reformulation of Equation~\ref{gbvisc} yields the following:

\begin{align}
    \eta &= \lim_{t_{s}\to\infty} \frac{V}{k_{B}T} \int_{0}^{t_{s}} \left < P_{ij}(t_{0})P_{ij}(t_{0}+t)\right >_{t_{0}}dt \\
    \eta &= \lim_{t_{s}\to\infty}\eta_{t_s} \\
    \eta_{t_s} &= \frac{V}{k_{B}T} \int_{0}^{t_{s}} \left < P_{ij}(t_{0})P_{ij}(t_{0}+t)\right >_{t_{0}}dt \label{gbvisc_finite} \\
    \eta_{t_s} &= \frac{V}{k_{B}T} \lim_{\delta t \to 0} \sum_{t=0}^{t_{s}} \delta t \left < P_{ij}(t_{0})P_{ij}(t_{0}+t)\right >_{t_{0}}
\end{align}

In Equation~\ref{gbvisc_finite}, $\eta_{t_{s}}$ is the approximate value of viscosity when the integral is only taken for a finite time $t_{s}$. The thermodynamic average $\left < \right>_{t_{0}}$ of pressure values was performed by averaging over several time origins, $t_{0}$, separated by a time lag. We take this average over different $t_{0}$ values, each separated by multiples of 0.1 $ps$, to obtain a value of $\eta_{t_{s}}$ every 100 $ps$. Pressure values with equal time separations are then averaged and integrated 100 times to get the total $\eta_{t_{s}}$ over the full 10 $ns$ isothermal hold. To further explain, consider the following for autocorrelation of a quantity $P=P_{ij}$:

\begin{align}
    \left < P(t_{0})P(t_{0}+t)\right > = C_{PP}(t_{0}, t_{0}+t) \\
    \nonumber \\
    \quad\text{(because of time-translation invariance)} \nonumber \\
    C_{PP}(t_{0}, t_{0}+t) = C_{PP}(j \Delta t) \\
    \nonumber \\
    \quad\text{(because of ergodicity)} \nonumber \\
    C_{PP}(j \Delta t) = \dfrac{1}{N-j}\sum_{i=0}^{N-1-j}P(i \Delta t)P((i+j)\Delta t)
\end{align}

\par
where $j$ is the separation between frames, $\Delta t$ is the sample interval, and $N$ is the total number of frames for a 100 $ps$ period. See Refs.~\cite{chemtext, gromacs} for further details. As an example, consider separations of $j \in \{0, 1, 2, N-1\}$ with $\Delta t = $ 100 frames (equivalently 0.1 $ps$):

\begin{align*}
    \quad\text{Average with zero lag at j = 0} \\
    C_{PP}(0) = \dfrac{1}{N}\sum_{i=0}^{N-1}P(100i)P(100i) \\ 
    \\
    \quad\text{Average with 100 frame lag at j = 1} \\
    C_{PP}(100) = \dfrac{1}{N-1}\sum_{i=0}^{N-2}P(100i)P((i+1)100) \\
    \\
    \quad\text{Average with 200 frame lag at j = 2} \\
    C_{PP}(200) = \dfrac{1}{N-2}\sum_{i=0}^{N-3}P(100i)P((i+2)100) \\
    \\
    \quad\text{Average with maximum lag at j = N-1} \\
    C_{PP}(100(N-1)) = P(0)P((N-1)100) \\
\end{align*}

\par
Each time lag from $C_{PP}(j \Delta t)$ was averaged. For example, the very first step in MD produces one autocorrelation measure of $P_{ij}$ with zero time lag. Then, the first 100 $ps$ interval generates another 1000 values that are averaged with the previous for a mean value from 1001 observations with zero time lag. After 200 $ps$, there are a total of 2001 values with zero time lag to average. The same procedure was repeated for each possible time lag for the total 10 $ns$ per isothermal hold. The integral with respect to each time lag average of the autocorrelation function of $P_{ij}$ was used to compute viscosity with Equation~\ref{final_visc} where $\tau = j\Delta t$.

\begin{equation}\label{final_visc}
    \eta_{t_{s}} = \frac{V}{k_{B}T}  \lim_{\delta \tau \to 0} \sum_{\tau=0}^{t_{s}} \delta \tau C_{PP}(\tau)
\end{equation}

\par
\begin{figure}[H]
\centering
\includegraphics[width=0.95\textwidth]{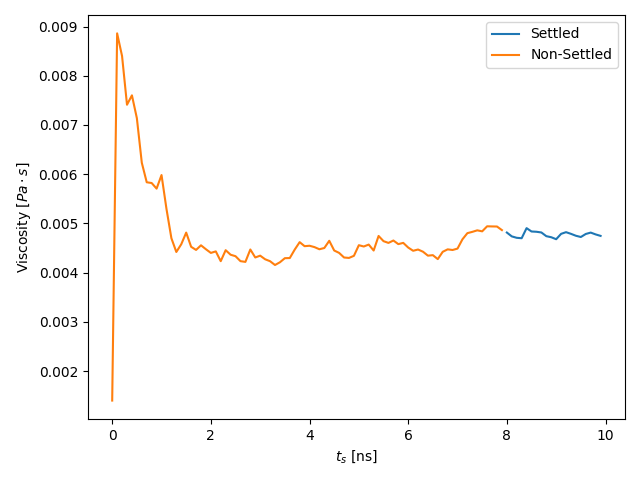}
\caption{The running integral for viscosity of a single run of $Cu_{50}Zr_{50}$ at 1100 K. The curve is the pressure autocorrelation integrals taken every of 100 $ps$. The blue portion of the curve contains data used for convergence analysis of viscosity.}
\label{visc}
\end{figure}

To ensure a settled viscosity measurement, the final 2 $ns$ were gathered from NVT simulations, the gradient of viscosity with respect to time was taken to acquire slopes, and then the mean of the slopes was taken. If the mean slope was below $10^{-5}$ $Pa$, then data were considered stable and therefore converged. For the converged cases, the average over the final 2 $ns$ was used to determine our viscosity measurement. We average values because there are some minor viscosity fluctuations as seen in Figure~\ref{visc}.

\section*{Appendix B}

\setcounter{equation}{0}
\renewcommand{\theequation}{B\arabic{equation}}

\setcounter{table}{0}
\renewcommand{\thetable}{B\arabic{table}}

\setcounter{figure}{0}
\renewcommand{\thefigure}{B\arabic{figure}}

\begin{table}[H]
\centering
\caption{The phase diagrams used from ASM International for $T_{l}$ values.}
\begin{tabular}{|l|l|}
\hline
{System} & {Reference}                                              \\ \hline
\hline
Ag-Cu            & Silver-Copper   Binary Phase Diagram (2007 Cao W.)              \\ \hline
Al-La            & Aluminum-Lanthanum Binary Phase Diagram   (2000 Okamoto H.)     \\ \hline
Al-Zr            & Aluminum-Zirconium Binary Phase Diagram   (2002 Okamoto H.) a   \\ \hline
Al-Cu            & Aluminum-Copper Binary Phase Diagram   (1991 Chen S.)           \\ \hline
Al-Ti            & Aluminum-Titanium Binary Phase Diagram   (2012 Wang H.)         \\ \hline
Al-Ni            & Aluminum-Nickel Binary Phase Diagram   (2005 Miettinen J.)      \\ \hline
Al-Sm            & Aluminum-Samarium Binary Phase Diagram   (2007 Delsante S.)     \\ \hline
Al-Co            & Aluminum-Cobalt Binary Phase Diagram   (2004 Ohtani H.)         \\ \hline
Cu-Zr            & Copper-Zirconium Binary Phase Diagram   (2010 Kang D.H.) a      \\ \hline
Fe-Ni            & Iron-Nickel Binary Phase Diagram (1991   Swartzendruber L.J.) d \\ \hline
Mg-Y             & Magnesium-Yttrium Binary Phase Diagram   (2008 Guo C.)          \\ \hline
Nb-Ni            & Niobium-Nickel Binary Phase Diagram (2007   Tokunaga T.)        \\ \hline
Ni-Ti            & Nickel-Titanium Binary Phase Diagram   (2010 Agraval P.G.)      \\ \hline
Ni-Zr            & Nickel-Zirconium Binary Phase Diagram   (2007 Wang N.)          \\ \hline
Pd-Si            & Palladium-Silicon Binary Phase Diagram   (2006 Du Z.) b         \\ \hline
\end{tabular}
\label{asm}
\end{table}

\printbibliography[heading=bibintoc]

\end{document}